# Exhaustive Exact String Matching: The Analysis of the Full Human Genome

Konstantinos F. Xylogiannopoulos
*Department of Computer Science*
*University of Calgary*
Calgary, AB, Canada
kostasfx@yahoo.gr

***Abstract*—Exact string matching has been a fundamental problem in computer science for decades because of many practical applications. Some are related to common procedures, such as searching in files and text editors, or, more recently, to more advanced problems such as pattern detection in Artificial Intelligence and Bioinformatics. Tens of algorithms and methodologies have been developed for pattern matching and several programming languages, packages, applications and online systems exist that can perform exact string matching in biological sequences. These techniques, however, are limited to searching for specific and predefined strings in a sequence. In this paper a novel methodology (called Ex2SM) is presented, which is a pipeline of execution of advanced data structures and algorithms, explicitly designed for text mining, that can detect every possible repeated string in multivariate biological sequences. In contrast to known algorithms in literature, the methodology presented here is string agnostic, i.e., it does not require an input string to search for it, rather it can detect every string that exists at least twice, regardless of its attributes such as length, frequency, alphabet, overlapping etc. The complexity of the problem solved and the potential of the proposed methodology is demonstrated with the experimental analysis performed on the entire human genome. More specifically, all repeated strings with a length of up to 50 characters have been detected, an achievement which is practically impossible using other algorithms due to the exponential number of possible permutations $\sum_{n=1}^{50} 4^n$ of such long strings.**

## I. Introduction

From the first developmental steps of computers and computer science, string matching and searching related problems became the most frequently studied. Some of the most important tasks assigned to computers are, for example, executing intensive and demanding calculations or performing searches in files and texts. More recently, with the explosion of data science and big data analytics, the desire for in depth pattern matching and detection is even more in demand. Machine learning and deep learning algorithms are used in computationally heavy problems with the aim of detecting patterns that can be used for further analysis and conclusion of meaningful and important attributes, for example, of time series, images etc.

One of the first problems studied in computer science is string matching in biological sequences. Analysis of biological sequences is considered to be a standard string-matching problem since all information in DNA, RNA and proteins is stored as nucleotide sequences using a simple and short alphabet. For example, in the DNA the whole genome is expressed with the four-letter based alphabet A, C, G and T representing Adenine, Cytosine, Guanine and Thymine respectively. What made the string matching problem important in bioinformatics from the very early stages of development was not just the obvious importance of biology related analysis of the life base information, but also the size of the strings that had to be analyzed, which was beyond any available computer capacity 20 or 30 years ago. Only the human genome consists of 24 chromosomes varying from 40 to 270 MB strings and a total size of approximately 3.4 GB. Although recent hardware improvements in memory and CPU have made an indirect brute force algorithm analysis possible, which could be extremely slow, more advanced techniques using special data structures such as suffix trees could still need memory sizes that are not easily available and can be achieved only with the use of clustering computing and systems such as Hadoop and Spark. For, example for the first human chromosome with size 270MB a suffix tree representation requires 26GB of memory [17].

Another example of the importance of string matching algorithms in bioinformatics is the new sequencing technology (NGS) used. The base of this new technology is the sequence alignment where very small fragments created by machines developed by companies such as Illumina, Roche, Life Technologies etc. have to be aligned to a previously well-known genome used as a reference [17]. What makes this task computationally difficult and also very important is not the single string matching alignment but the millions of fragments which have to be aligned.

Apart from string alignment, a more complicated example in bioinformatics data analytics could also be the comparison of genomes between humans or between different species. For example, a redesign of species classification based on DNA information could be attempted. In such a case, instead of the standard human 3.4 GB string, a multiple of millions of times would be needed for the comparison of millions of DNA sequences among all species, making such a task practically impossible. Furthermore, even if hardware is available, algorithms would have to process an enormous amount of data making the task impossible for brute force or inefficient algorithmic approaches.

In this paper a novel and advanced methodology will be described, which combines state-of-the-art data structures and algorithms specifically designed for pattern detection and text mining. It will be presented how it is possible to achieve extraordinary results by creating a pipeline that optimizes the Longest Expected Repeated Pattern Reduced Suffix Array (LERP-RSA) data structure, designed and developed for text mining, and the All Repeated Pattern Detection (ARPaD) algorithm. In brief, LERP-RSA is a variation of the standard Suffix Array with the significant difference of using the actual suffix strings, while ARPaD is an algorithm, both recursive and non-recursive, with the ability to detect every pattern that occurs at least twice in a string. LERP-RSA and ARPaD have been exhaustively used in the past to detect all repeated patterns that exist in big strings using standard commodity computers and they could be used to address biological strings of terabytes. For example, in [22], a string composed from the first 1 Trillion digits of π has been analyzed and more than 426 billion repeated patterns have been discovered. The results of the analysis include the two longest repeated patterns of length 23 that exist in the first 1 Trillion digits of π, which can be verified using the new Google pi-api [32]. Furthermore, the major advantage of this methodology compared to others is that the algorithm is pattern agnostic, which means that it takes no input. According to literature, exact string matching algorithms take as an input the string that needs to be discovered and they return its occurrences and their positions. Instead of this, ARPaD is executed without requiring any input and detects every repeated pattern that exists in a string regardless of its attributes, i.e., frequency, length, overlapping, alphabet etc. Moreover, an optimized method for repeated pattern detection with application in bioinformatics will be presented, a problem which has specific characteristics that makes such an analysis even harder.

The rest of the paper is organized as follows: Section II presents related work in string matching and more specifically exact string matching. Section III defines the problem and gives the motivation behind it. Section IV presents the proposed methodology for all repeated pattern detection in biological sequences. Section V describes the experiments conducted with the available dataset of the full human genome and discusses the corresponding results. Finally, Section VI presents the conclusions and future extensions of the presented work.

## II. Related Work

Pattern detection in biological sequences, such as DNA, is a special case of the string matching problem, which has attracted huge attention since the 70's. Several studies and surveys have been conducted in the past years which examine most of the newly presented algorithms [15, 16, 17]. Based on these surveys, the problem and the algorithms developed can be classified into two main categories, exact matching and approximate matching [15, 17]. In the first one we search to find strings, or patterns, that exist in a sequence and absolutely match the input string. In the second one we want to find the best possible approximation, which can be an outcome of deletion, insertion or mutation. Exact pattern matching algorithms are of particular interest since we perform not just single or multiple pattern matching, but we detect every possible recurring pattern. A further classification of exact string matching algorithms, based on the general methodology followed, can be described as below.

The first class is the indexed or character based approach [15, 16, 17]. This class includes standard brute force algorithms where each character is compared with the reference string. This, of course, increases the computational time complexity of the algorithm regardless of the absence of any kind of preprocessing of the dataset. Other approaches that fall into this category are the well-known Boyer-Moore algorithm, the Morris-Pratt algorithm etc. [2, 15, 17] The most famous algorithm of this category though is the Knuth-Morris-Pratt algorithm designed approximately 40 years ago [1, 16, 17]. This algorithm uses a supplementary table which encompasses shift information that describes by how many characters the pattern should be moved when a character fails to match the running character in the string [1, 16, 17].

The Boyer-Moore algorithm is a standard for string matching, usually used as a benchmark reference [2, 15, 17]. The basic characteristic of this algorithm is the shifting step during which the corresponding shift table provides information regarding the number of characters that can be avoided when a mismatch occurs [15]. Furthermore, it includes a skip table in addition to the shift table of the KMP algorithm [15]. The aforementioned algorithm, apart from being used as a reference, has also been used as a basis for several variations, extensions and improvements [15, 17]. First of all, Boyer-Moore-Smith proposed the BMS algorithm which computes the move with the character. Despite the possible problems in some cases with the rightmost character, it usually behaves well [23]. Another extension is the Apostolico-Giancarlo algorithm based on both KMP and BM algorithms [3]. Basically, this algorithm keeps track of the pattern that was matched successfully and although it accesses each character twice at most it has performance problems with very long patterns [3]. Other variations are the Raita algorithm, which is based on dependencies that exist between consecutive characters [29] and Crochermore et al. who developed the turbo BM algorithm based on dynamic simulation [4]. A more recent algorithm of this category is the BBQ algorithm by Ahmad, which uses two parallel pointers to search from opposite directions (left-right) [5]. Furthermore, in order to achieve better performance, more advanced, hybrid approaches have been used such as the KMPBS algorithm by Xian [6] and the Cao et al. which uses statistical probability [7].

The second major class of exact pattern matching algorithms is the hashed-based [15, 16, 17]. This family of algorithms is based on calculating hashing values for the character to match rather than performing exact matching to the character as the abovementioned algorithms. This technique can significantly improve calculation time since it uses integer values for comparison instead of characters [24]. Yet, as is common with hashing, this method suffers from the hashing collision problem when two dissimilar strings are mapped on the same hashing integer [15]. One of the first algorithms was the Karp-Rabin algorithm [25] which was used for string matching problems and it is based on modular arithmetic for hashing. Another algorithm, which uses a q-gram approach is the Lecroq algorithm [26]. In this case the sequence is divided into *n* subsequences and then the subsequences are used for matching. Non q-gram algorithms have also been developed where the full searching pattern is encoded. Such algorithms are the Wu and Manber [27] or the Faro [28].

Another class is the suffix automata-based algorithms. These algorithms comprise two "distinct automata constructors: deterministic acyclic finite state automaton […] and suffix automaton […] for matching." [15, pp.8]. The most famous algorithm of this class is the Knuth-Morris-Pratt algorithm [1]. The concept behind this algorithm is that during the left to right scanning, a decision is made regarding the number of patterns to be shifted in order to skip redundancy when a mismatch occurs [1, 15]. Other Directed Acyclic Word Graph-based algorithms (DAWG) are the Backward non-deterministic algorithm based on the non-deterministic automaton approach [8, 16] and the Double-forward DAWG algorithm which uses two automata [9, 16]. Furthermore, additional algorithms have been developed recently such as the multi-window integer comparison algorithm based on suffix string from Hongbo et al. [11, 15], the Franek-Jennings-Smyth string matching algorithm [12], the automata skipping algorithm developed by Waga et al. [10] and more.

Additionally, several hybrid approaches exist which perform better in many cases since they combine good characteristics from different methodologies. Such algorithms are for example Navarro's algorithm [13] which can bypass characters using suffix automaton and Wuu's algorithm [14] which is an improvement of KMP algorithm that uses a suffix tree for string matching that allows us to move shifting both left-to-right and top-to-bottom.

Concluding, for bioinformatics specific purposes, there are several other algorithms used, for example, by the National Center for Biotechnology Information (NCBI). The most current algorithm used by NCBI is the Basic Local Alignment Search Tool (BLAST) and its variants [31]. The BLAST algorithm is used for comparing basic sequences, such as nucleotides sequences, found in DNA and/or RNA. The algorithm takes as inputs the desired sequence to search for and the sequence to search against. It is important to mention that BLAST can also perform inexact string matching. Another algorithm is the Smith-Waterman algorithm [30], which although more accurate than BLAST, it is considerably slower and, thus, not used for large genome datasets.

## III. Problem Definition

So far, we have described many algorithms which can be classified in many different categories based on their core characteristics. However, all string-matching algorithms have a common parameter: they use an input string that they try to match on the search string and detect every possible occurrence. Yet, there are patterns whose existence we are not aware of but they could be very important because, for example, they may reoccur in different genes related to a disease and a possible similar medical treatment could be available.

In that case we would like to have a methodology that can detect every reoccurring pattern regardless of (a) the number of occurrences, (b) position in a chromosome, (c) positions in different chromosomes or (d) having occurrences between different humans or other species. In the aforementioned cases, because of the enormous amount of data and possible patterns, the algorithm should not only be accurate but it also has to be extremely efficient in order to perform the analysis in a meaningful time span. Furthermore, the analysis should be executed on normal hardware with minimum resources in order to (a) keep cost low and (b) allow scale up with more advanced resources for larger datasets.

## IV. Methodology

The methodology which will be described here is a combination of the Longest Expected Repeated Pattern Reduced Suffix Array (LERP-RSA) data structure [20, 21, 22], the Moving LERP [20, 22] and the All Repeated Pattern Detection (ARPaD) algorithms [21, 22]. Several components have been put together to create an advanced pipeline that can be used to perform exhaustive string matching, i.e., detect every repeated string exist in a biological sequence.

### *A. LERP-RSA Data Structure*

The LERP-RSA data structure is a text mining special purpose data structure which is designed and developed to work with the ARPaD algorithm. LERP-RSA is a variation of the suffix array as described by Manber and Myers [19]. A suffix array is the array of the indexes of the lexicographically sorted suffixes of a string. However, LERP-RSA, instead of using the indexes of the suffix strings, uses the actual suffixes. Anyone could claim that this has quadratic space complexity, yet, with the use of the Probabilistic Existence of Longest Expected Repeated Pattern Theorem [21, 22], the complexity can be downgraded to log-linear with regard to the input string. The Theorem briefly states that:

**Theorem:** If a string is considerably long and random and a pattern is reasonably long then the probability that the pattern repeats in the string is extremely small.

Therefore, we can truncate the length of the suffix strings significantly by using the following Lemma [21, 22]:

**Lemma:** Let $S$ be a random string of size $n$, constructed from a finite alphabet $\Sigma$ of size $m \geq 2$, and an upper bound of the probability $P(X)$ is $\overline{P(X)}$, where $X$ the event "LERP is the longest pattern that occurs at least twice in $S$." An upper bound for the length $l$ of the Longest Expected Repeated Pattern (LERP) we can have with probability $P(X)$ is:

$$\bar{l} = \left\lceil log_m \frac{n^2}{2\overline{P(X)}} \right\rceil$$

where $l \ll n$ and $\overline{P(X)} > 0$.

The Lemma is directly concluded from the Theorem and it has been proven in [21, 22]. The advantage of the Lemma comparing to other methods, e.g., constructing the corresponding suffix tree and determining the longest repeated pattern on it, is that we simply have to execute in advance a numerical calculation on the Lemma formula and no other preprocess is required.

For example, if the string that we want to examine for repeated patterns is *CATTATTAGGA* then the suffix strings are presented in "Fig. 1.a". When we lexicographically sort the suffix strings then we conclude with the array of "Fig. 1.b". The classic suffix array is the indexes of the lexicographically sorted suffix strings in "Fig. 1.b". However, the LERP-RSA is presented in "Fig. 1.c" where the suffix strings are truncated to have length at most 4 characters since we do not expect to have reoccurring patterns with longer length. The overall time and space complexity of the LERP-RSA is $O(nlogn)$.

The LERP-RSA data structure has some unique features that makes it a state-of-the-art data structure for text mining purposes. The most important attributes are (a) classification based on the alphabet, (b) network and cloud distribution based on the classes, (c) full and semi parallelism, (d) self-compression and (e) indeterminacy. These attributes will be described further in the next paragraphs and it will be shown how in combination with ARPaD algorithm it can achieve extraordinary results. However, in order for LERP reduction to work properly the Theorem requires that a string should be random. In general, this means that every character of the alphabet occurs with the same frequency and this property should be valid for reasonably long substrings, according to Calude's Theorem [18]. For many cases that LERP-RSA has been used so far, this property is valid and the data structure has behaved accordingly. However, biological sequences do not have this randomness property. This is expected because in biological sequences each nucleotide should have a purpose to appear at certain positions. Furthermore, many long, overlapping, single nucleotide sub-sequences exist. This deviation from Theorem requirements makes LERP-RSA application harder but it will be explained in the next paragraphs how this can be solved.

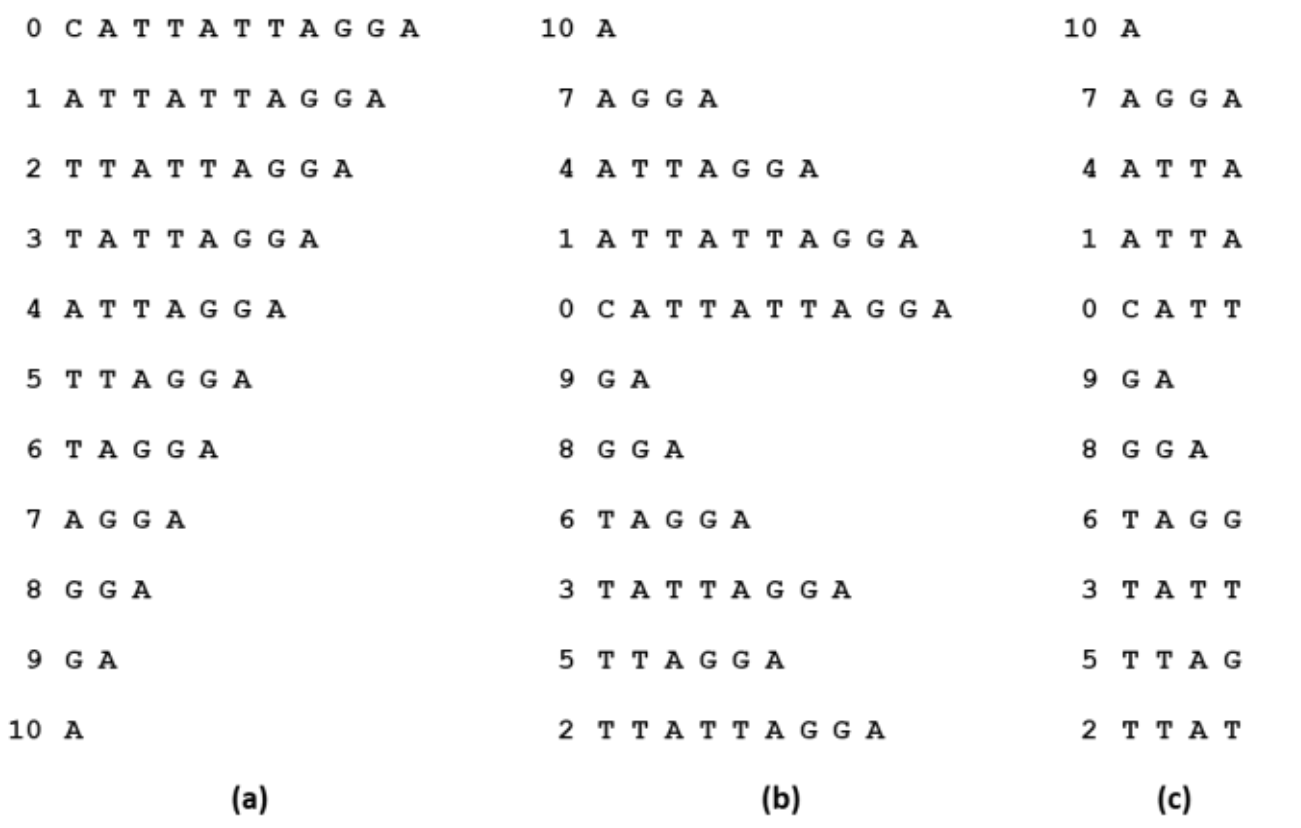

Fig. 1. Suffix Array and LERP-RSA for *CATTATTAGGA*

## B. *ARPaD Algorithm*

When the LERP-RSA data structure is completed then we use the All Repeated Patterns Detection (ARPaD) algorithm. The algorithm has two versions, the recursive left-to-right and the non-recursive top-to-bottom [22]. Both versions have the same time complexity $O(nlogn)$. The recursive works as follows. It starts with the first letter of the alphabet, *A*, and counts how many strings starts with it "Fig. 2.a". Since more than one occurrence exists then pattern *A* is a repeated pattern. The algorithm constructs a longer pattern with the first letter *A*, the *AA*. This does not exist and the algorithm continues with the other letters of the alphabet until it finds twice the pattern *AT* "Fig. 2.b". The process is repeated for longer patterns, starting with *ATA*, until it finds *ATT* "Fig. 2.c" and *ATTA* "Fig. 2.d" which also occur twice. When it finishes with letter *A* it continues and repeats until it covers all alphabet letters. At the end ARPaD algorithm has discovered all repeated patterns "Fig. 2.d". The non-recursive top-to-bottom version works in a similar way by comparing directly suffix string tuples.

As we can observe from the ARPaD execution example, the algorithm works on each alphabet letter separately. This allows us to use one of the best properties of LERP-RSA, which is classification by splitting our data structure to smaller partitions based on the alphabet, for example, by creating classes with all suffix strings starting with *A*, *C*, *G* and *T*. Then we can execute ARPaD in parallel or semi-parallel mode on those classes and reduce the execution time to a quarter at best, if we assume equidistribution of the classes. Yet, in biological sequences the equidistribution is not granted since we know that *A* and *T* have approximately the double frequency of *C* and *G*. However, this is not a problem because we can create even more classes using different Classification Level for some or all classes. For example, if we assume that class *A* holds 50% of the suffix strings and the other four letters share the same frequency then we can create for class *A* subclasses of Classification Level 2, i.e., *AA*, *AC*, *AG*, *AT*. Doing this we will have seven classes, i.e., *AA*, *AC*, *AG*, *AT*, *C*, *G* and *T*, with approximately the same frequency and we can execute ARPaD in parallel over these seven classes. Here we can use another important attribute of LERP-RSA since ARPaD is executed independently on each class. Because of this, we can use the network and/or cloud distribution property which allows to use completely isolated and diversified hardware to analyze each class instead of using expensive hardware infrastructure or clustering frameworks such Hadoop and Spark. Therefore, we can create and store classes on different locations and execute ARPaD locally or remotely on them and collect the results when the execution is completed.

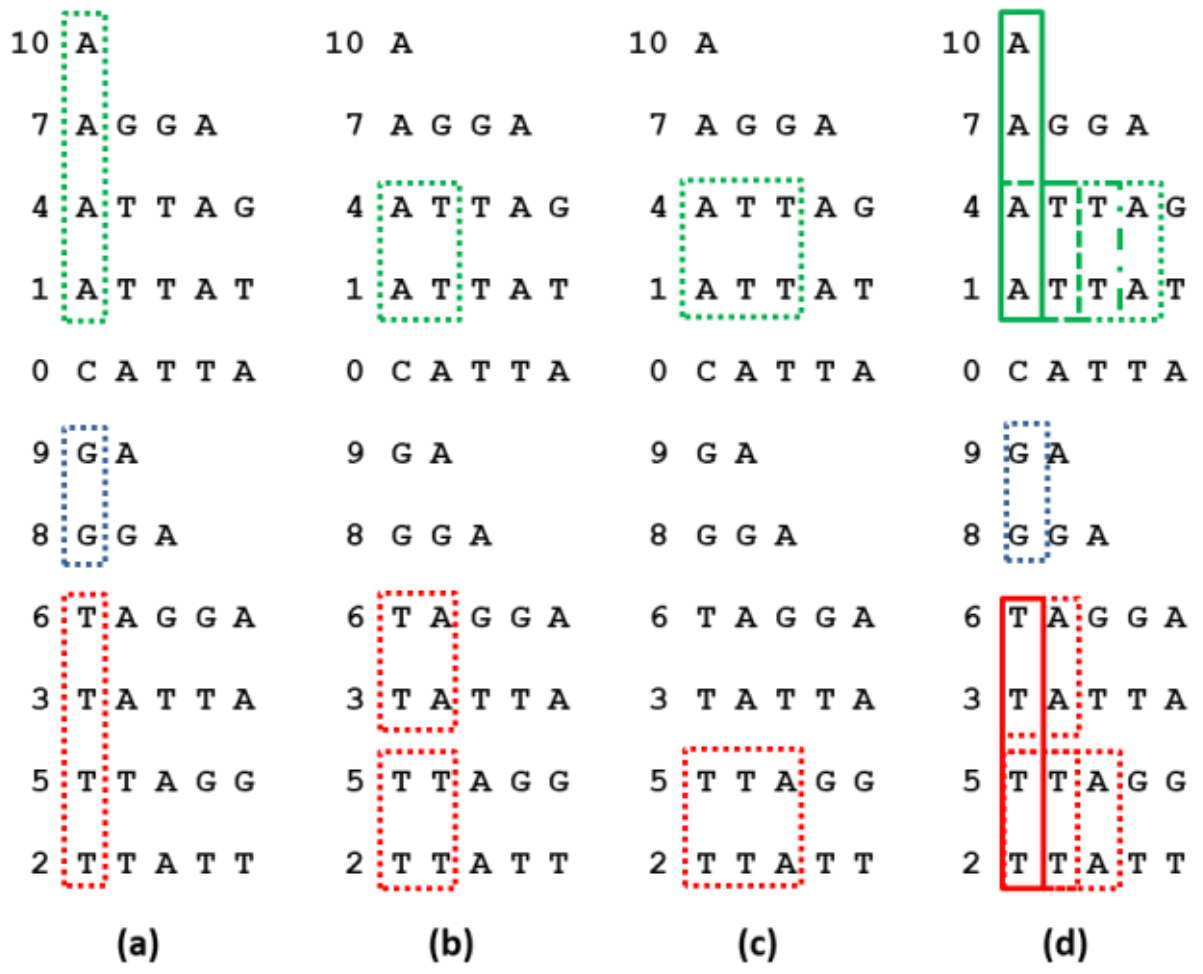

Fig. 2. ARPaD execution example on *CATTATTAGGA*

## C. *Multivariate LERP-RSA Data Structure*

A core characteristic of biological sequences is their complexity in size and number. For example, the human genome has 24 chromosomes and if we need to compare DNA sequences among many different individuals the chromosomes number under investigation can rise significantly. Therefore, we need data structures and algorithms that are capable searching and finding patterns effectively in multiple biological sequences. This can be achieved with the Multivariate LERP-RSA.

When we have more than two strings to analyze, regardless of the dependency, we can create a joined LERP-RSA data structure as follows. First, we create the LERP-RSA for each string, either a single data structure or with multiple classes using different Classification Levels. Then we merge these separate LERP-RSA or classes. Since the data structures are already sorted, the merge process is very simple and fast, having an overall time and space complexity $O(nlogn)$. At the end we create a new LERP-RSA data structure which holds not only the suffix strings and their positions but also the string where these occur. For example, in "Fig. 3." Two strings have been used namely *CATTATTAGGA* and *CATTCA*. The bold font numbers in front of each suffix string represent the string index while the normal font number represents the positions. After creating the LERP-RSA for each string, these two data structures are merged in one “Fig. 3.a”. After that ARPaD is executed as described in the previous section and discovers all repeated patterns that exist in this new LERP-RSA. It is important to mention that with this process we can detect (a) patterns that exist only in one string, as if we have executed ARPaD on a single string, for example, *ATTA* at position 0.1 and 0.4, (b) patterns that exist in both strings such as *T* at positions 0.2, 0.3, 0.5, 0.6, 1.2 and 1.3 or (c) patterns that do not repeat in each string but exist as repeated patterns only because of the specific combination of both strings such as *CATT* at position 0.0 and 1.0.

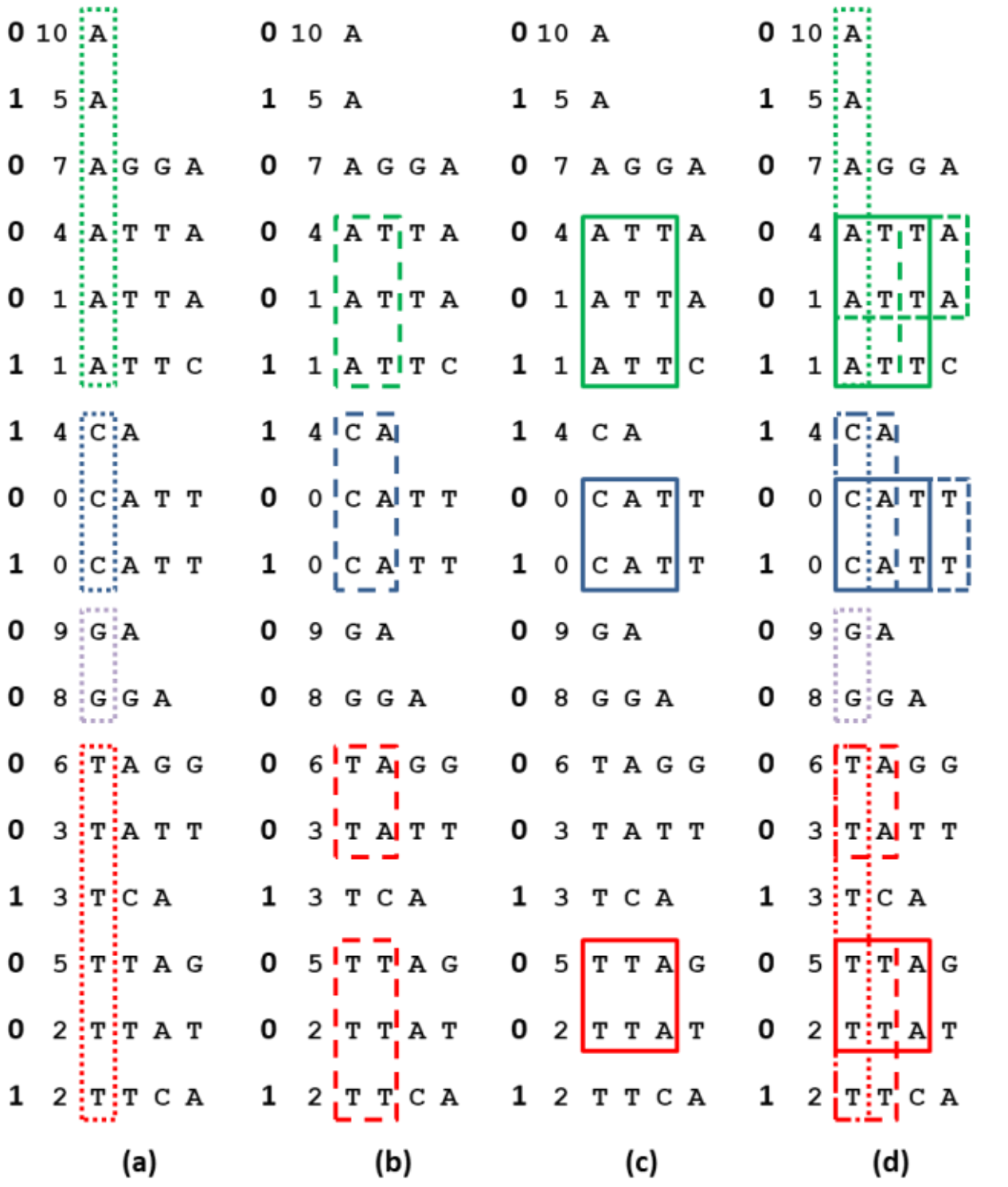

Fig. 3. Multivariate LERP-RSA for *CATTATTAGGA* and *CATTCA*

## D. *MLERP-ARPaD Algorithm*

As was described in Section IV.B, biological sequences do not have the randomness property and, therefore, it is expected that LERP Theorem will not provide the appropriate value for the LERP-RSA construction. In order to bypass this problem, we can use the Moving LERP algorithm, specifically design to solve this problem. The MLERP works as follows: First, the LERP value is calculated based on the string attributes, i.e., length and alphabet size, and the LERP-RSA is constructed using this value. After that ARPaD is executed and discovers every repeated pattern with a length up to LERP value. For the patterns that have been discovered and have length exactly LERP we can distinguish two cases, (a) the pattern has length exactly LERP or (b) the pattern is longer with a length greater than LERP. Since it is impossible to know in advance the hidden length, we create a new LERP-RSA data structure with LERP value double the original used from the Theorem. This new LERP-RSA though is created only for suffix strings that exist at the positions that the patterns with a length LERP have been discovered. After that we have a new, significantly smaller, data structure and we execute the ARPaD algorithm again but with the use of the Shorter Pattern Length (SPL) which has the value of the original LERP. The SPL allows us to avoid executing ARPaD for lengths smaller or equal to original LERP and wasting time for rediscovering patterns with a length up to original LERP. Additionally, we avoid duplicates in our results which would require an extra process at the execution end to clean them. If patterns with a length double LERP are discovered we repeat the process with new LERP and SPL (equal to previous LERP) until we have no more repeated patterns.

In order to maximize the performance of the MLERP algorithm and fully utilize our available hardware it is preferable instead of doubling the LERP value in each round to increase LERP according to available resources. For example, if after the first round 20% of the discovered patterns have length exactly LERP then we can use as new LERP value five times the original LERP value and create a new data structure with exactly the same required space. This way we can usually reduce the repetition rounds to no more than three.

An example of MLERP algorithm can be observed in “Fig. 4”. In this example the initial value for LERP is three “Fig. 4.a”. After ARPaD execution three patterns with a length exactly 3 have been discovered namely *ATT* at positions 0.4, 0.1 and 1.1, *CAT* at positions 0.0 and 1.0 and *TTA* at position 0.5 and 0.2. For these positions only a new LERP-RSA is constructed in the second round of MLERP execution with LERP value 6 “Fig. 4.b”. Then ARPaD is executed again and discovers the longest patterns *ATTA* and *CATT* with a length four. Since the longest repeated pattern has length smaller than the new LERP, MLERP execution terminates. As we can observe from the example the patterns detected are exactly the same as in the first example of section IV.B.

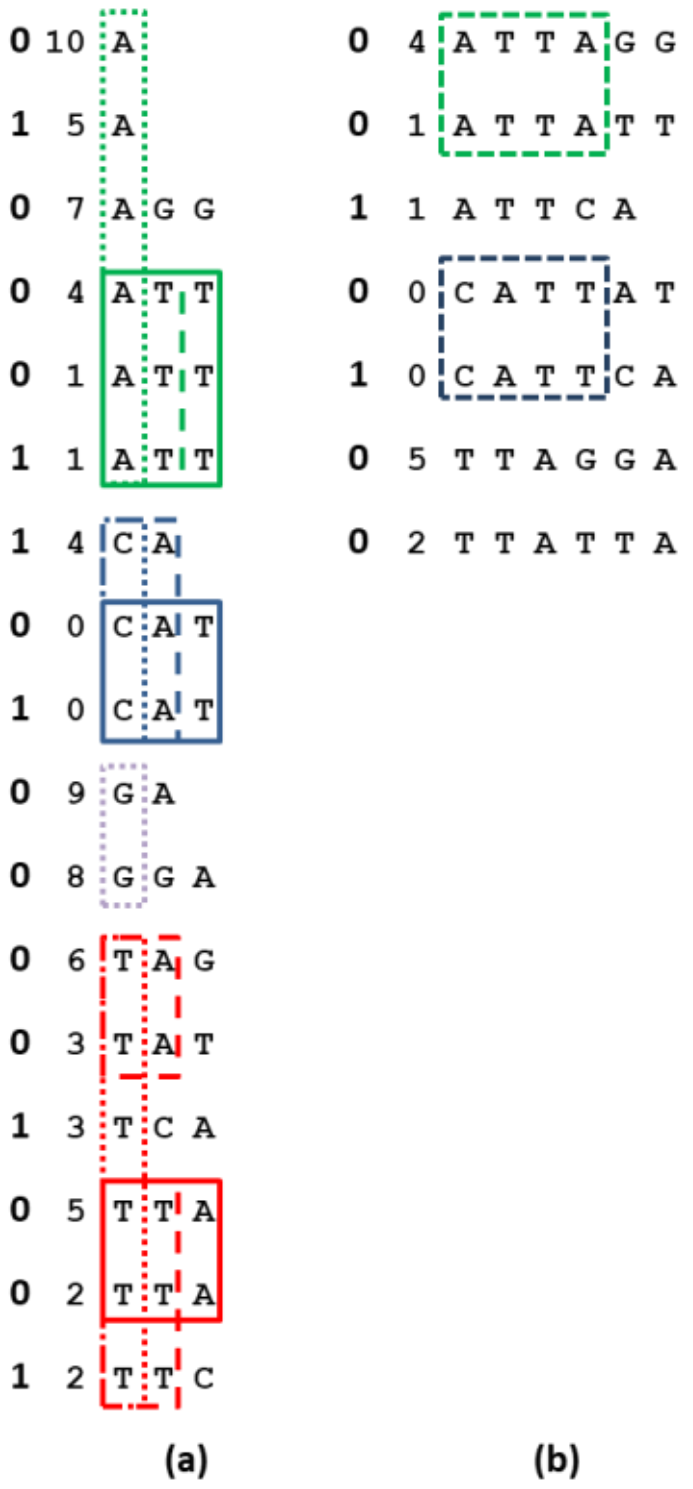

Fig. 4. MLERP algorithm example for initial LERP value 3

## E. *Metadata Analyses*

After the completion of the data analysis and pattern detection on multiple strings, several metadata analyses can be performed. These analyses depend on several factors such as comparing areas of different chromosomes from the same genome, comparing genes among different individuals, different species etc. The importance of the full analysis and repeated patterns detection is that we need to execute our further, detailed, meta analyses in the results already calculated, which make this process extremely easy and fast. For example, in most cases of simple string matching we just need a simple binary search with time complexity $O(logn)$. Furthermore, the knowledge acquired from such analyses could enlighten scientists by providing information about unknown properties of biological sequences. For example, a common pattern could be discovered in different genes related to different diseases that could help to provide common treatment among them.

### F. *Exhaustive Exact String Matching (Ex2SM)*

Concluding, we need to describe the full process for the exhaustive exact string matching process. This is a pipeline as described in the following figure "Fig. 5" and it was thoroughly presented in the previous sections. Briefly it can be formulated as follows.

1) Construct the LERP-RSA for each string (1..n) and class (1..m)
2) Merge LERP-RSA from all strings per class (1..m)
3) Execute ARPaD on Multivariate/Multivariable LERP-RSA for each class (1..m)
4) If patterns with a length equal to LERP have been detected then execute MLERP by setting SPL equal to LERP, increasing LERP and repeating from step (1) for positions belonging to patterns found with length exactly LERP, otherwise continue
5) Perform metadata analyses

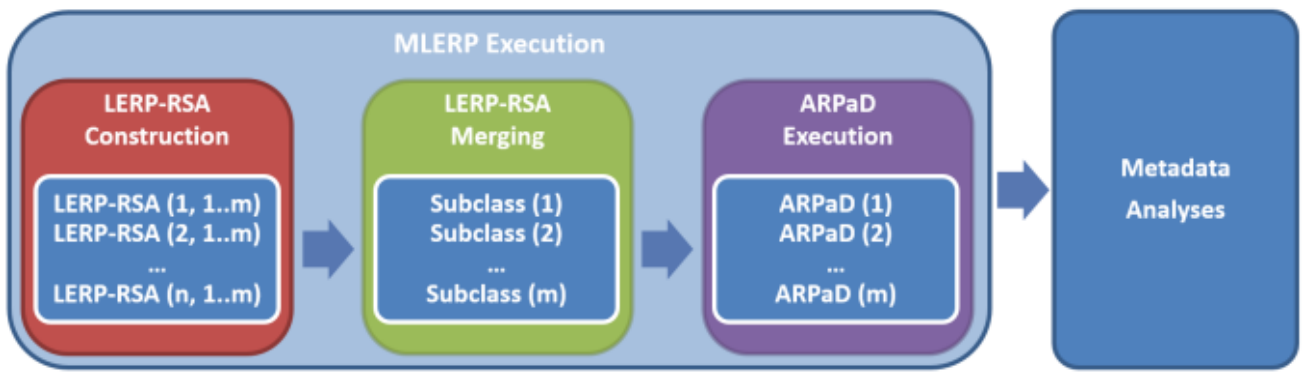

Fig. 5. Exhaustive Exact Pattern Matching Pipeline (Ex2SM)

The overall space and time complexity of the methodology presented here is $O(nlogn)$. In Table I we can observe that each component of the pipeline has loglinear time complexity except the construction of the LERP-RSA which is linear, while the space complexity for the several LERP-RSA phases is also loglinear. It is important to mention that because of the network/cloud distribution property of the LERP-RSA the time and space complexity could be practically linear $O(n)$, with regard to the input string, since we can construct classes with Classification Level $logn$.

TABLE I. Ex2SM Pipeline Time and Space Complexity

| Component | Time Complexity | Space Complexity |
|---|---|---|
| LERP-RSA Construction | $O(n)$ | $O(nlogn)$ |
| LERP-RSA Sorting | $O(nlogn)$ | $O(nlogn)$ |
| LERP-RSA Merging | $O(nlogn)$ | $O(nlogn)$ |
| MLERP | $O(nlogn)$ | $O(nlogn)$ |
| ARPaD | $O(nlogn)$ | - |
| Overall | $O(nlogn)$ | $O(nlogn)$ |

## V. Experimental Analysis

For the experimental analysis of the Exhaustive Exact String Matching (Ex2SM) methodology the human genome GRCh38.p12 has been used, found on the National Center for Biotechnology Information (NCBI) [28] in its FASTA format. This consists of 24 strings representing the corresponding chromosomes and forming a total dataset of approximately 3.4 GB. The strings have been cleaned from irrelevant characters such as *N*, *W* etc. and headers that exist at several positions inside chromosomes. For the analysis a laptop with an Intel i7 CPU at 2.6 GHz has been used with 16 GB RAM and an external disk of 500 GB. The results of this analysis are enormous and for practical reasons only few, interesting, metadata analyses will be presented here.

TABLE II. Patterns Detected

| Pattern Length | Total Patterns | Total Occurrences | Cumulative Patterns |
|---|---|---|---|
| 2 | 16 | 3,091,632,200 | 16 |
| 3 | 64 | 3,091,630,801 | 80 |
| 4 | 256 | 3,091,629,408 | 336 |
| 5 | 1,024 | 3,091,628,025 | 1,360 |
| 6 | 4,096 | 3,091,626,646 | 5,456 |
| 7 | 16,384 | 3,091,625,272 | 21,840 |
| 8 | 65,536 | 3,091,623,903 | 87,376 |
| 9 | 262,144 | 3,091,622,536 | 349,520 |
| 10 | 1,048,576 | 3,091,621,173 | 1,398,096 |
| 11 | 4,190,700 | 3,091,617,150 | 5,588,796 |
| 12 | 16,354,255 | 3,091,358,953 | 21,943,051 |
| 13 | 58,073,353 | 3,087,325,380 | 80,016,404 |
| 14 | 165,037,816 | 3,053,472,437 | 245,054,220 |
| 15 | 376,760,490 | 2,919,960,491 | 621,814,710 |
| 16 | 520,247,161 | 2,474,854,256 | 1,142,061,871 |
| 17 | 427,936,880 | 1,825,473,229 | 1,569,998,751 |
| 18 | 290,086,195 | 1,350,042,758 | 1,860,084,946 |
| 19 | 208,279,847 | 1,099,444,223 | 2,068,364,793 |
| 20 | 171,422,042 | 976,933,300 | 2,239,786,835 |
| 21 | 156,055,756 | 910,867,680 | 2,395,842,591 |
| 22 | 149,389,034 | 868,019,798 | 2,545,231,625 |
| 23 | 146,071,688 | 835,117,116 | 2,691,303,313 |
| 24 | 144,062,368 | 807,092,451 | 2,835,365,681 |
| 25 | 142,611,988 | 781,998,960 | 2,977,977,669 |
| 26 | 141,425,630 | 758,960,438 | 3,119,403,299 |
| 27 | 140,372,312 | 737,553,895 | 3,259,775,611 |
| 28 | 139,419,197 | 717,557,253 | 3,399,194,808 |
| 29 | 138,546,059 | 698,778,006 | 3,537,740,867 |
| 30 | 137,724,362 | 681,066,964 | 3,675,465,229 |
| 31 | 136,933,862 | 664,311,143 | 3,812,399,091 |
| 32 | 136,164,173 | 648,412,247 | 3,948,563,264 |
| 33 | 135,409,024 | 633,295,371 | 4,083,972,288 |
| 34 | 134,665,078 | 618,918,717 | 4,218,637,366 |
| 35 | 133,927,583 | 605,234,394 | 4,352,564,949 |
| 36 | 133,193,832 | 592,207,337 | 4,485,758,781 |
| 37 | 132,463,714 | 579,801,656 | 4,618,222,495 |
| 38 | 131,740,136 | 567,978,403 | 4,749,962,631 |
| 39 | 131,022,257 | 556,705,424 | 4,880,984,888 |
| 40 | 130,311,648 | 545,957,260 | 5,011,296,536 |
| 41 | 129,614,517 | 535,708,565 | 5,140,911,053 |
| 42 | 128,928,750 | 525,929,911 | 5,269,839,803 |
| 43 | 128,249,388 | 516,597,170 | 5,398,089,191 |
| 44 | 127,574,921 | 507,683,425 | 5,525,664,112 |
| 45 | 126,903,783 | 499,167,698 | 5,652,567,895 |
| 46 | 126,238,460 | 491,033,033 | 5,778,806,355 |
| 47 | 125,575,740 | 483,262,678 | 5,904,382,095 |
| 48 | 124,916,776 | 475,846,639 | 6,029,298,871 |
| 49 | 124,269,519 | 468,775,047 | 6,153,568,390 |
| 50 | 123,630,166 | 462,026,886 | 6,277,198,556 |

For the full analysis of the DNA, Classification Level 2 has been used to create the LERP-RSA. This means that first the 16 classes for suffix strings starting with *AA, AC, …, TG, TT* have been constructed for each chromosome and then these partial LERP-RSA data structures have been merged to the full LERP-RSA of each class. The initial LERP value that has been used is 25 since this is the approximate optimal length the Lemma calculates. For this LERP approximately 142 million patterns have been detected with a total number of occurrences close to 782 million and cumulative patterns with length up to 25 approximately 3 billion (Table II). Based on these results a second round has been run with MLERP and new LERP value 50. In this case approximately 123 million patterns have been detected with a total number of occurrences approximately 462 million. In total, more than 6.3 billion patterns have been detected with a length from 2 up to 50. What is important from this preliminary analysis is that there is a significant number of very long repeated patterns because of the non-random behavior of the DNA.

Another very important outcome of the result is related to the patterns that have a length exactly 50. By performing a metanalysis per class, i.e., *AA, AC, …, TG, TT,* it has been observed that for all classes we can find repeated patterns of length 50 with a similar number of occurrences in every chromosome. However, for class *TG* there are repeated patterns only in chromosomes 2, 4, 7, 9, 15, 16, 18 and 20 “Fig. 6”. Furthermore, patterns belonging to the *TG* class are significantly more comparing to the other classes.

Another important finding of the metanalysis that has been conducted is that for the majority of patterns with a length 50 the repetitions is limited to just 2 occurrences (Table III) representing approximately 70% of the occurrences.

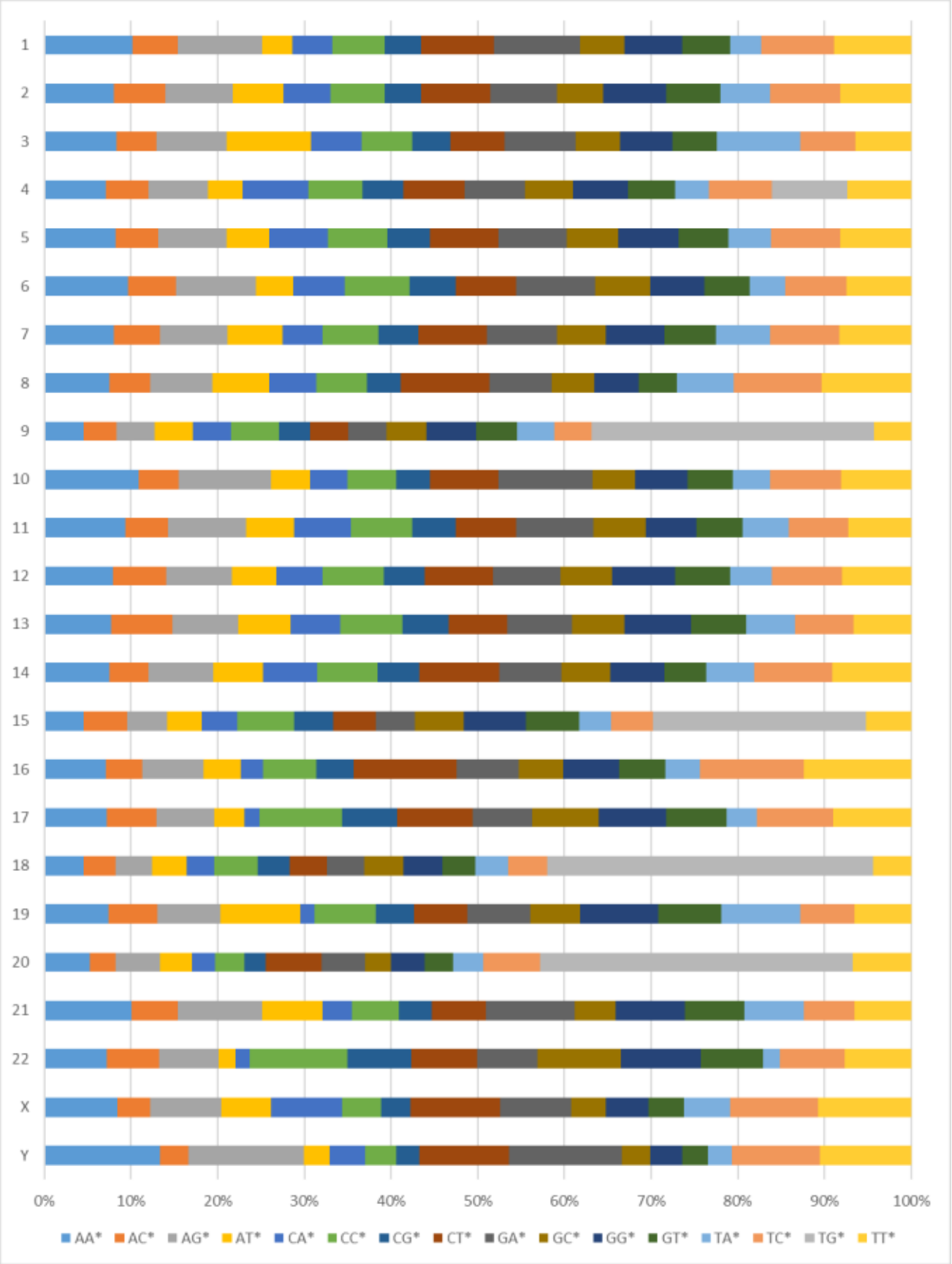

Fig. 6. Patterns frequency with a length 50 per chromosome and class

TABLE III. OCCURRENCES PER CLASS

| Class | Patterns with 2 Occurrences | Percentage of Patterns with 2 Occurrences | Patterns with 2 Occurrences at the Same Chromosome | Percentage of Patterns Occurring in same Chromosome |
|---|---|---|---|---|
| AA | 7,899,543 | 72.62% | 616,980 | 7.81% |
| AC | 4,452,290 | 71.03% | 367,314 | 8.25% |
| AG | 6,237,160 | 70.77% | 500,181 | 8.02% |
| AT | 6,267,578 | 72.59% | 507,452 | 8.10% |
| CA | 6,497,774 | 70.96% | 529,535 | 8.15% |
| CC | 5,058,266 | 70.34% | 410,357 | 8.11% |
| CG | 1,113,590 | 67.18% | 94,062 | 8.45% |
| CT | 6,264,874 | 70.89% | 502,880 | 8.03% |
| GA | 5,286,709 | 70.43% | 439,941 | 8.32% |
| GC | 4,131,152 | 70.02% | 324,563 | 7.86% |
| GG | 5,106,568 | 69.74% | 420,431 | 8.23% |
| GT | 4,494,404 | 70.81% | 372,150 | 8.28% |
| TA | 5,163,247 | 73.27% | 408,181 | 7.91% |
| TC | 5,299,061 | 70.69% | 438,622 | 8.28% |
| TG | 6,574,333 | 70.74% | 539,992 | 8.21% |
| TT | 8,089,179 | 71.66% | 631,525 | 7.81% |

TABLE IV. MOST OCCURRED PATTERN WITH A LENGTH 50

| Pattern | Total Occurrences |
|---|---|
| AAGAAAGAAAGAAAGAAAGAAAGAAAGAAAGAA AGAAAGAAAGAAAGAAA | 8,993 |
| ACTGCAAGCTCCGCCTCCCGGGTTCACGCCATTCTC CTGCCTCAGCCTCC | 5,600 |
| AGAAAGAAAGAAAGAAAGAAAGAAAGAAAGAAA GAAAGAAAGAAAGAAAG | 8,716 |
| ATATATATATATATATATATATATATATATATATAT ATATATATATATAT | 6,013 |
| CATAGTATTCCATGGTGTATATGTGCCACATTTTCT TAATCCAGTCTATC | 5,716 |
| CCTGTAGTCCCAGCTACTCGGGAGGCTGAGGCAGG AGAATGGCGTGAACC | 7,131 |
| CGCCTGTAGTCCCAGCTACTCGGGAGGCTGAGGCA GGAGAATGGCGTGAA | 4,978 |
| CTTTCTTTCTTTCTTTCTTTCTTTCTTTCTTTCTTTCT TTCTTTCTTTCT | 8,685 |
| GAAAGAAAGAAAGAAAGAAAGAAAGAAAGAAAG AAAGAAAGAAAGAAAGA | 8,784 |
| GCCTGTAGTCCCAGCTACTCGGGAGGCTGAGGCAG GAGAATGGCGTGAAC | 6,082 |
| GGTTCACGCCATTCTCCTGCCTCAGCCTCCCGAGTA GCTGGGACTACAGG | 7,258 |
| GTTCACGCCATTCTCCTGCCTCAGCCTCCCGAGTAG CTGGGACTACAGGC | 6,136 |
| TATATATATATATATATATATATATATATATATATA TATATATATATATA | 5,818 |
| TCTTTCTTTCTTTCTTTCTTTCTTTCTTTCTTTCTTTC TTTCTTTCTTTC | 8,741 |
| TGAAAAAGGAAATATCTTCCCATAAAAACTAGACA GAAGCATTCTCAGAA | 6,010 |
| TTTCTTTCTTTCTTTCTTTCTTTCTTTCTTTCTTTCTT TCTTTCTTTCTT | 8,954 |

Moreover, in Table III we can observe that for the patterns that occur only twice more than 90% occurs in different chromosomes.

In Table IV we can observe the patterns with a length 50 that have the most occurrences at each class. The reason of presenting the specific length is to present the complexity of the problem and the strength of the proposed methodology. If we need to discover all strings having length just 50 characters using any other algorithm, we will need to check for $4^{50}$ or $2^{100}$ string arrangements, since we cannot know in advance which string exists or not. This is practically impossible with a brute force attack technique for any algorithm. If we assume that searching for each string arrangement in a sequence needs one millisecond, the full process will require approximately 32 trillion years, using a single core. Even with the use of super computers and billion cores, GPUs etc. such analysis at this time is practically infeasible.

## VI. Conclusions

In the current paper a new methodology (Ex2SM) has been presented that allows exhaustive exact string matching with application in biological sequences for the detection of all repeated exact strings. The methodology is built as a pipeline of different components, i.e., data structures and algorithms, that altogether can achieve extraordinary results. As a proof of concept, the analysis of the full human genome has been executed and more than 6.2 billion patterns with a length of up to 50 characters have been detected, having approximately a total number of 70 billion occurrences. The presented methodology is string agnostic, i.e., it does not require an input string in order to search for it. Moreover, it can be executed on multiple sequences, e.g., chromosomes, and detect patterns that exist as repeated either in a single chromosome or only among different chromosomes. The overall time and space complexity is $O(nlogn)$ and in contrast to other methods it can perform an exhaustive exact string matching to detect every repeated string in a few hours using commodity hardware.

Although many string matching algorithms exist, the problem of detecting all repeated strings for bioinformatics purposes was not fully addressed in literature. In future work a more detailed and thorough analysis will be conducted presenting several metanalyses of interest in biology, medicine etc. A comparative genomic study among several species, including human, will also take place. Moreover, an inexact pattern matching methodology using LERP-RSA and ARPaD will be presented.

## VII. References

[1] Knuth D.E., Morris J.H., Pratt V.R. (1977). “Fast pattern matching in strings.” SIAM Journal on Computing, 6(2), pp. 323-350

[2] Boyer, R. S. and Moore, J. S. (1977). “A fast string searching algorithm.” Communications of the ACM, pp. 762-772

[3] Apostolico, A. and Giancarlo, R. (1986) “The Boyer-Moore-Galil String Searching Strategies Revisited.” (in English), SIAM Journal on Computing, 15(1), pp. 98-105

[4] Crochemore, M., Czumaj, A., Gasieniec, L., Jarominek, S., Lecroq, T., Plandowski, W., Rytter, W., (1994) “Speeding up two string-matching algorithms.” Algorithmica, pp. 247-267

[5] Ahmad, M. K. (2014) “An Enhanced Boye-Moore Algorithm (Doctoral dissertation).” Middle East University

[6] Xian-Feng, H., Yu-Bao, Y., Xia, L. (2010) “Hybrid pattern-matching algorithm based on BM-KMP algorithm.” 3rd International Conference In Advanced Computer Theory and Engineering (ICACTE), (5), pp. 310

[7] Cao, Z., Zhenzhen, Y., Lihua, L. (2015) “A fast string matching algorithm based on lowlight characters in the pattern.” 7th International Conference on Advanced Computational Intelligence (ICACI), pp. 179-182

[8] Commentz-Walter, B. (1979). “A string matching algorithm fast on the average.” Springer, pp. 118-132

[9] Allauzen, R. (2000). “Simple optimal string matching algorithm.” Algorithms, pp. 102-116

[10] Masaki, W., Hasuo, I., Suenag, K. (2017) “Efficient online timed pattern matching by automata-based skipping.” International Conference on Formal Modeling and Analysis of Timed Systems, Springer, pp. 224-243

[11] Hongbo, F., Shupeng, S., Jing, Z., Li., D. (2015) “Suffix Type String Matching Algorithms Based on Multi-windows and Integer Comparison." In International Conference on Information and Communications Security, Springer, pp. 414-420

[12] Franek, F. J., Jennings, C.G., Smyth, W.F. (2007) “A simple fast hybrid pattern matching algorithm." Journal of Discrete Algorithms, pp. 682-695

[13] Navarro, G. (2001) “NR-grep: a fast and flexible pattern-matching tool.” Softw., Pract. Exper., 31, 1265-1312

[14] Lu, H. T. and Yang, W. (2001) “A simple tree pattern-matching algorithm.” in Proceedings of the Workshop on Algorithms and Theory of Computation

[15] Hakak, S., Kamsin, A., Shivakumara, P., Gilkar, G. A., Khan, W. Z., Imran, M. (2017) “Exact String Matching Algorithms: Survey, Issues and Future Reseach Directions”. Preparation of Papers for IEEE Transcations and Journals, 2017

[16] Faro, S. (2016). “Evaluation and Improvement of Fast Algorithms for Exact Matching on Genome Sequences.” AlCoB.

[17] Chen, Y. (2018). “String Matching in DNA Databases”, Open Access Biostatistics and Bioinformatics, 1(4)

[18] Calude, C., (1995) “What is a Random String?” Journal of Universal Science, 1(1), pp. 48–66

[19] Manber, U. and Myers, G., (1990) “Suffix arrays: a new method for on-line string searches.” Proceedings of the First Annual ACM-SIAM Symposium on Discrete Algorithms, pp. 319–327

[20] Xylogiannopoulos, K. F., Karampelas, P., Alhajj, R. (2014) “Analyzing very large time series using suffix arrays” Appl. Intell., 41(3), pp.941–955

[21] Xylogiannopoulos, K. F., Karampelas, P., Alhajj, R. (2016) “Repeated patterns detection in big data using classification and parallelism on LERP reduced suffix arrays” Appl. Intell., 45(3), pp. 567– 597

[22] Xylogiannopoulos, K. F., (2017) “Data structures, algorithms and applications for big data analytics: single, multiple and all repeated patterns detection in discrete sequences.” PhD thesis

[23] Smith, P.D. (1991) “Experiments with a Very Fast Substring Search Algorithm.” Softw., Pract. Exper., 21, 1065-1074

[24] AbdulRazzaq, A. A., Rashid, N. A. A., Hasan, A. A., Abu-Hashem, M. A, “The exact string matching algorithms efficiency review.” Global Journal on Technology, pp. 576-589, 2013.

[25] Karp, R. M. and Rabin, M. O. (1987) “Efficient Randomized Pattern-Matching Algorithms.” IBM Journal of Research and Development, 31(2), pp. 249-260

[26] Lecroq, T. (2007) “Fast exact string matching algorithms.” Information Processing Letters, 102(6), pp. 229-235

[27] Wu, S. and Manber, U. (1994) “A fast algorithm for multi-pattern searching.” Department of Computer Science, University of Arizona, Tucson, AZ, Report TR-94-17

[28] National Center for Biotechnology Information (NCBI) ftp://ftp.ncbi.nlm.nih.gov/genomes/Homo_sapiens/

[29] Raita, T. (1992) “Tuning the Boyer-Moore-Horspool string searching algorithm.” Software: Practice and Experience, pp. 879-884

[30] Smith, T. F. and Waterman, M. S. (1981) "Identification of Common Molecular Subsequences" Journal of Molecular Biology. 147 (1): 195–197

[31] National Center for Biotechnology Information (NCBI) https://blast.ncbi.nlm.nih.gov/Blast.cgi

[32] Google pi-api https://pi.delivery/